\documentclass[runningheads]{llncs}

\usepackage{cite}
\usepackage{float}
\usepackage{bibcheck}
\usepackage{graphicx}
\usepackage{amssymb}
\usepackage{amsopn}
\usepackage{amstext}
\usepackage{lscape}
\usepackage{multirow}
\usepackage[justification=centering]{caption}
\usepackage[justification=centering]{subcaption}
\usepackage[backref]{hyperref}

\newcommand{\reffig}[1]{Figure \ref{#1}}
\newcommand{\reftab}[1]{Table \ref{#1}}
\newcommand{\refsec}[1]{Section \ref{#1}}

\begin{document}
	\title{Attention-based Graph ResNet for Motor Intent Detection from Raw EEG signals}
	\titlerunning{Attention-based Graph ResNet for Motor Intent Detection}
	\author{Shuyue Jia, Yimin Hou, Yan Shi, Yang Li}
	\institute{Northeast Electric Power University, China}
	\maketitle
	
	\begin{abstract}
	In previous studies, decoding electroencephalography (EEG) signals has not considered the topological relationship of EEG electrodes. However, the latest neuroscience has suggested brain network connectivity. Thus, the exhibited interaction between EEG channels might not be appropriately measured via Euclidean distance. To fill the gap, an attention-based graph residual network, a novel structure of Graph Convolutional Neural Network (GCN), was presented to detect human motor intents from raw EEG signals, where the topological structure of EEG electrodes was built as a graph. Meanwhile, deep residual learning with a full-attention architecture was introduced to address the degradation problem concerning deeper networks in raw EEG motor imagery (MI) data. Individual variability, the critical and longstanding challenge underlying EEG signals, has been successfully handled with the state-of-the-art performance, 98.08\% accuracy at the subject level, 94.28\% for 20 subjects. Numerical results were promising that the implementation of the graph-structured topology was superior to decode raw EEG data. The innovative deep learning approach was expected to entail a universal method towards both neuroscience research and real-world EEG-based practical applications, e.g., seizure prediction.
	
	\end{abstract}
	
	\section{Introduction}
	The recent decade has witnessed a flourishing market in the brain-computer interface (BCI) to assist and rehabilitate patients from strokes, Parkinson's disease, and brain injuries\cite{biasiucci2018brain, shanechi2019brain}. Electroencephalography (EEG) obtains growth in popularity because of its non-invasiveness, high temporal resolution, and portability\cite{shanechi2019brain}. In particular, the EEG-based motor imagery (MI) technique has been extensively employed to manipulate the peripherals via neural activities\cite{mahmood2019fully}. De facto, individual variability poses a critical and longstanding challenge for current EEG signal processing. Hence, it is eagerly expected for researchers to handle such a concern to better understand the mechanism of cognitive functions and build robust EEG-based BCI systems. 
	
	Traditional works analyzed EEG data without taking consideration of the topological relationship of EEG electrodes\cite{mahmood2019fully, hou2020novel, dose2018end}. The latest neuroscience, however, has suggested brain dynamic functional connectivity\cite{lv2014holistic, lv2017n, vidaurre2017brain}. Thus, the exhibited interaction of EEG channels might not be well reflected via Euclidean distance. The Graph Convolutional Neural Network (GCN) has generated widespread research interest, which has proven superior to process and analyze graph-structured data. Spectral GCN was primarily studied since it well defined a localized operator for convolutions on graphs, and managed to process graph signals\cite{shuman2013emerging}. Reference\cite{defferrard2016convolutional} proposed an effective and efficient GCN approach by constructing fast localized graph filters. A few works have applied such a base model to classify EEG tasks, primarily for emotion recognition\cite{zhang2019gcb, song2018eeg, wang2018eeg}. However, there remains a vacancy to employ the GCN approach in the area of EEG MI. Therefore, this paper presented a novel structure of the GCN, an Attention-based graph ResNet, to achieve precise detection of human motor intentions. The main contributions were precisely summarized: (1) To date, this was the first work to detect human motor intents from raw EEG signals via the GCN approach. (2) The intrinsic topological relationship of EEG electrodes was built as a graph, which has proven superior to classify and analyze EEG signals. (3) This paper entailed a novel approach intending to address the universal EEG-based problems in neuroscience, and pave the road to build practical clinical applications.
	
	\section{Methods}
		\subsection{Graph Convolutional Neural Network}\label{Graph Convolutional Neural Network}
			\subsubsection{Graph Convolution}\label{Graph Convolution}
			An undirected graph is represented by the Laplacian matrix $L$, which contains the graph weights and degrees. The normalized graph Laplacian is defined as $L=I_{N}-D^{-1/2}AD^{-1/2}$, where $A$ is an adjacency matrix describing graph weights, $D$ is a degree matrix to represent node degrees. We used Absolute Pearson's Matrix $P$, which is the absolute of Pearson's Matrix as the adjacency matrix to measure the correlations between nodes. The degree matrix is measured as $D_{ii}=\sum_{j=1}^{N} A_{ij}$. Since the graph Laplacian is a real symmetric and positive semidefinite matrix, it can be factored via the graph Fourier transform as $L=U\Lambda U^{T}$, in which $U=\left[u_{0}, \ldots, u_{N-1}\right]$, and $\Lambda=\operatorname{diag}\left(\left[\lambda_{0}, \ldots, \lambda_{N-1}\right]\right)$ are the eigenvectors and eigenvalues, respectively. Graph convolution is defined:
			\begin{equation}
				y=U g_{\theta}(\Lambda) U^{T} x 
			\end{equation}  
			Where the Chebyshev polynomial filter $g_{\theta}(\Lambda)=\sum_{k=0}^{K-1} \theta_{k} T_{k}(\Lambda)$ is the most frequently-used filter $g_{\theta}$. $\theta \in \mathbb{R}^{K}$ is the model parameters, and $T_{k}(\Lambda)$ is the $K^{th}$ order Chebyshev polynomial approximation. The convolutional operation on the graph is then defined as:
			\begin{equation}
				y=\sum_{k=0}^{K-1} \theta_{k} T_{k}(L) x
			\end{equation}
			
			\subsubsection{Graph Pooling}\label{Graph Pooling}
			Defferrard \emph{et al.} proposed a fast graph pooling approach, including graph coarsening and balanced binary tree one-dimensional pooling, which has proven efficient\cite{defferrard2016convolutional}. A balanced binary tree is put forward to store the unarranged graph nodes. Based on the Graclus multilevel clustering algorithm, the normalized cut is minimized via the greedy method to coarsen the nodes\cite{dhillon2007weighted}, and then it max-pooled the nodes in the binary tree. In this way, the number of nodes will be divided into two, and the graph pooling operation is accomplished\cite{defferrard2016convolutional}. 
		
		\subsection{Residual Learning and Attention Mechanism}\label{Residual Learning and Attention Mechanism}
		The performance of DL models almost reached a plateau with a growing depth of networks. The phenomenon is well-known as the network degradation. Such a concern can be tackled via the residual learning framework, by which the networks can still converge and achieve higher accuracy when the number of layers increased\cite{he2016deep, li2019deepgcns}. In this work, residual learning was introduced to explore the performance of deeper networks for raw EEG signals. Previously, a stack of layers approximates the underlying mapping, which is from input $x$ to output $H(x)$. In contrast, with regard to residual learning, the layers fit the residual function $F(x)=H(x)-x$, where $x$ is the input, and $H(x)=F(x)+x$ is the originally desired mapping\cite{he2016deep}. As a result, the output of a stack of layers is $F(x)+x$, and it can be implemented via the shortcut connection.
		
		Discovered from the human vision, researchers have attracted considerable interest in the attention mechanism\cite{vaswani2017attention}. Previous works have either recognized equal contribution regarding the decoding tasks for all the EEG electrodes or selected the most influential electrodes (e.g., from the brain motor cortex). The principle of the attention mechanism is demonstrated as follows. 
		\begin{equation}
			u=\tanh \left(w y+b\right)
		\end{equation}
		\begin{equation}
			\alpha=\frac{\exp \left(u^{\top} u_{w}\right)}{\sum \exp \left(u^{\top} u_{w}\right)}
		\end{equation}
		\begin{equation}
			s=\alpha y
		\end{equation}
		In which $y$ is the original output of each layer, $u$ is a FC layer (In the experiment, we used tanh activation function), $\alpha$ is a Softmax layer, and $s$ is the attention-based final output. $w$, $u_{w}$, and $b$ are trainable weights and biases. The weighted method selects the most contributed features with regard to the classification task\cite{yang2016hierarchical}. A full-attention architecture was presented to preserve the details from raw EEG signals, and select the most crucial info. Each layer of the graph ResNet is followed and connected with an attention layer. And all the attention layers are jointly trained with the network.  
		
		\subsection{Attention-based Graph ResNet}\label{Attention-based Graph ResNet}
		The introduced approach was a novel structure of the GCN. It designed with the deep residual learning, and each layer (i.e., graph convolution, graph pooling, and fully-connected layer) was followed by an individual attention mechanism, which has formulated a full attention architecture. The residual learning framework makes it possible for deeper models to converge. The attention mechanism assigns weights to EEG electrodes, and selects and sticks out the most significant features to enhance the capability of the presented model. With regard to the EEG Motor Movement/Imagery Dataset, it is an international 10-10 system with 64 EEG electrodes. Thus, the dimension of the input, i.e., graph Laplacian, was 64$\times$64, followed by 12 layers of graph convolutions, and each two of them connected with a graph max-pooling layer, which reduced the dimension by two. The order of Chebyshev polynomial approximation for graph convolutional filters we used in our experiment was three. The activation function of graph convolution was Leaky ReLu, which was used after the batch normalization. The cross-entropy with L2 norm (0.001 regularization parameter) was applied as the loss function. We utilized the Adam Optimizer with 0.001 and 0.0001 learning rate for the graph-level and subject-level training, respectively. The framework was implemented via TensorFlow 1.14.0 under NVIDIA GTX1060 6G GPU and CUDA toolkit 10.0. 
		
	\section{Numerical Experiments}\label{Numerical Experiments}
		\subsection{Description of Dataset}\label{Description of Dataset}
		In the experiment, the EEG Motor Movement/Imagery Dataset (Physionet database) was used to validate the model. It performed four MI tasks, including imaging opening and closing left fist (denoting L), right fist (denoting R), both fists (denoting B), and both feet (denoting F)\cite{goldberger2000physiobank}. The sample rate was 160 Hz, and we used a four-second experimental duration, i.e., 640 time points per trial, to analyze the data\cite{hou2020novel}. 109 participants ($S_1-S_{109}$) took part in the experiment with 84 trials (4 tasks$\times$7 trials$\times$3 runs) for every participant. The detailed description of the dataset can be found in our previous work\cite{hou2020novel}. In this paper, the number of samples was 53,760 per subject. During all the experiments, the total dataset was randomly divided into ten pieces. Nine of them were selected as the training set, and the last one was regarded as the test set.
		
		\subsection{Model Comparison}\label{Model Comparison}
\begin{figure}[h]
			\centering
			\begin{minipage}[t]{.48\linewidth}
				\includegraphics[width=2.5in]{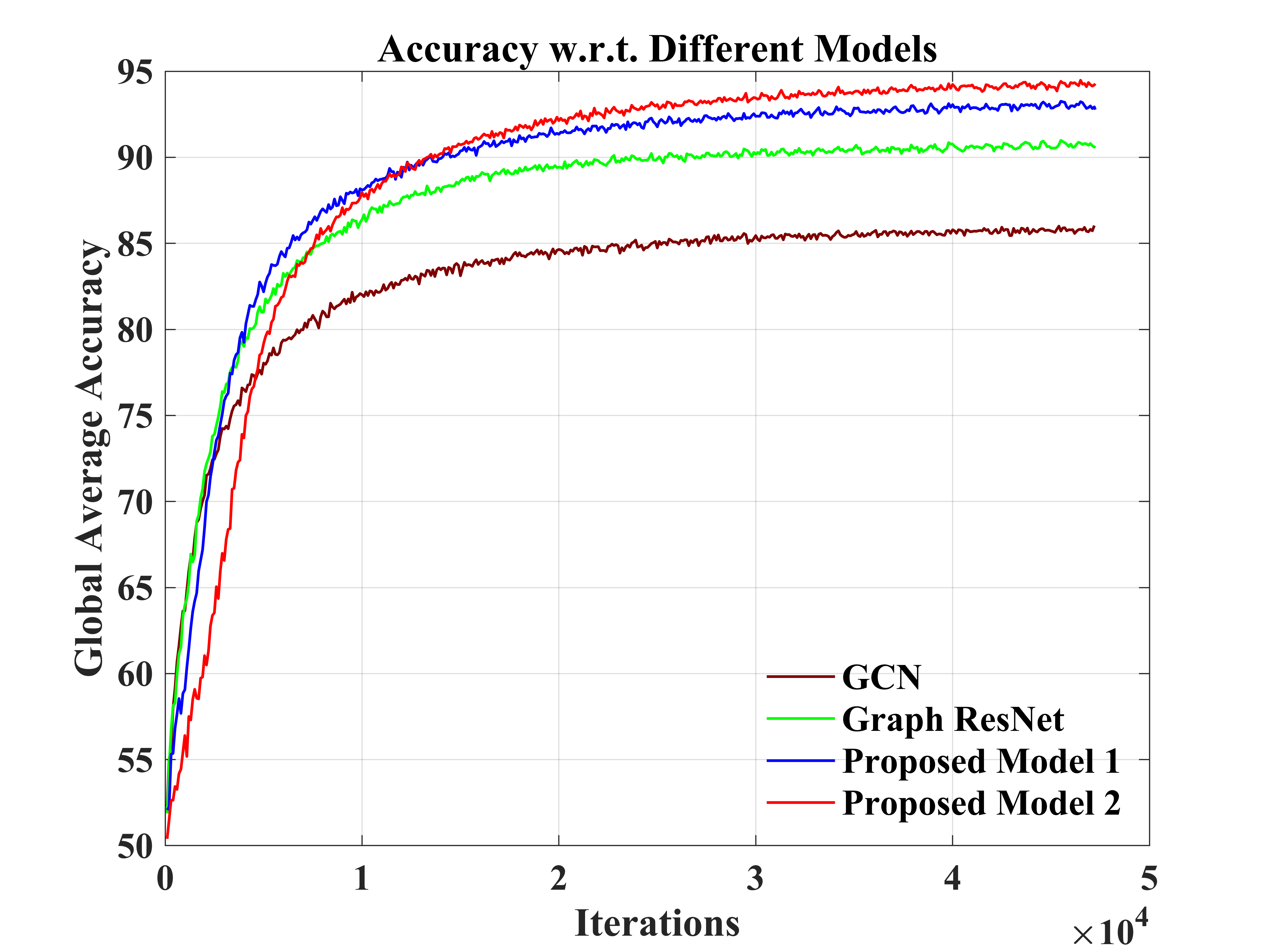}
				\subcaption{Accuracy of different models over iterations}
				\label{Accuracy_Model_Comparison}
			\end{minipage}
			\begin{minipage}[t]{.48\linewidth}
				\includegraphics[width=2.5in]{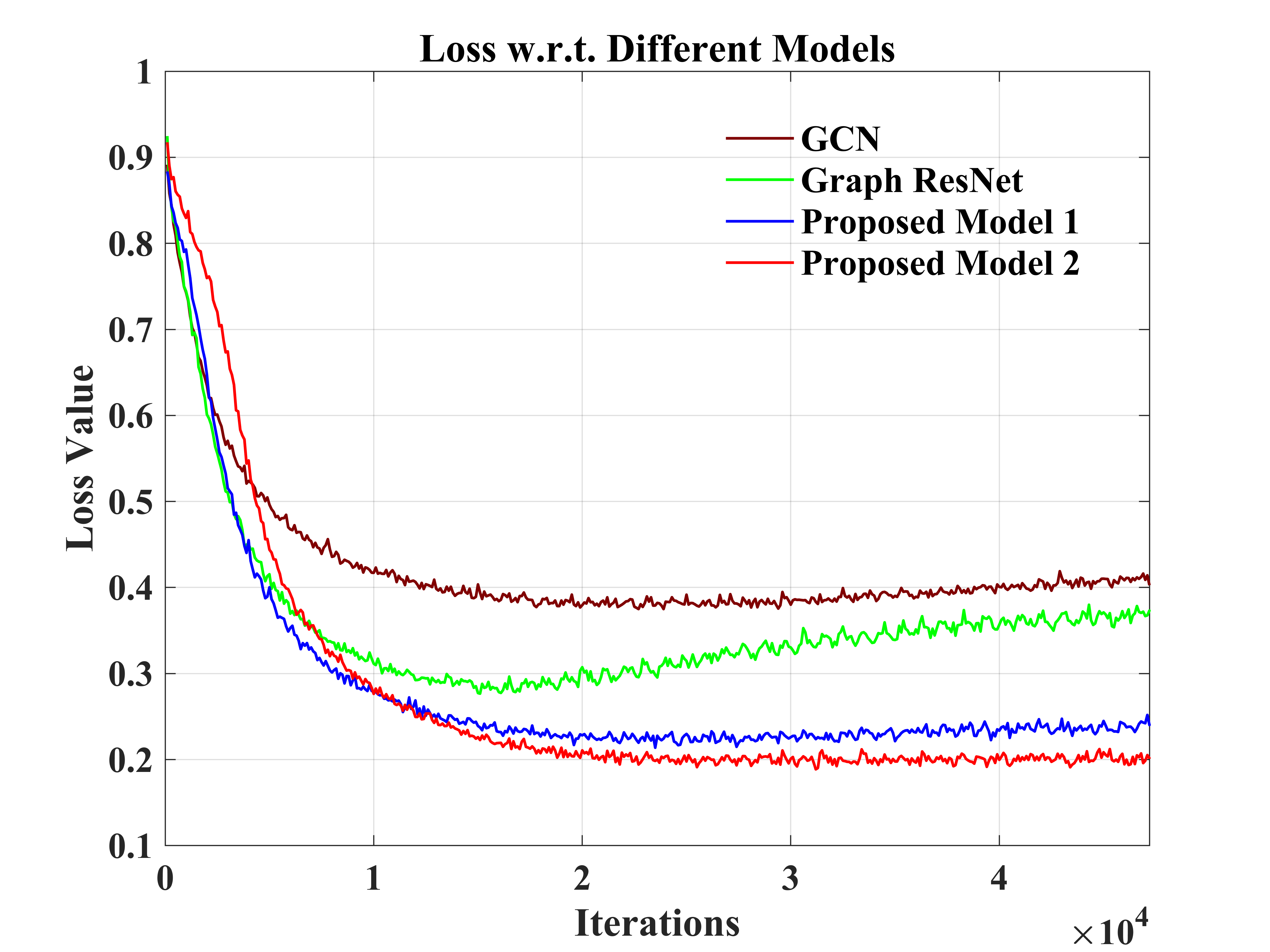}
				\subcaption{Loss value of different models over iterations}
				\label{Loss_Model_Comparison}
			\end{minipage}
			\caption{Different models' performance comparison (accuracy and loss value) over iterations.}
			\label{Accuracy_Loss_Model_Comparison}
		\end{figure}

		We compared the performance of various models, including the GCN, graph ResNet, Attention-based graph ResNet with two-layer (Proposed Model 1), and three-layer (Proposed Model 2) graph conv in the residual block. The dataset collected from 20 subjects ($S_1-S_{20}$) was utilized. Thus, the number of samples were 1,075,200 (20 subjects$\times$53,760 samples). The accuracy and loss over iterations curves were demonstrated in\reffig{Accuracy_Loss_Model_Comparison}. 
		
		\reffig{Accuracy_Model_Comparison} has shown the classification accuracy of different models. It was evident that all the GCN-based approach has achieved competitive results ($>$85\% accuracy). The reason for this phenomenon was that this work has taken consideration of the topological relationship of EEG electrodes. Compared with the GCN approach (86.03\% accuracy), the graph ResNet model increased the classification accuracy by 4.63\% via the deep residual learning. It, however, should also be admitted that as illustrated in \reffig{Loss_Model_Comparison}, the overfitting had occured as the model got complicated. Fortunately, after applying the attention mechanism, the problem has been effectively suppressed. Furthermore, the Attention-based graph ResNet has achieved dominant results at the group-level (20 subjects), which was supreme to decode EEG MI tasks. The proposed method 2 achieved 94.28\% accuracy, which was a little higher than the proposed method 1 (92.77\%). This was because there were more graph convolutions, thus the features can be extractd comprehensively, and the inter-subject variability can be filtered in a superior way. In other words, the residual learning approach promoted the classification performance compared with the original GCN model since it can tackle the problem of network degradation. Plus, the attention mechanism enhanced the performance again by assigning weights to the graph ResNet layer, which selected and highlighted the essential features in the layer. 
	 	
	 	\subsection{Subject-level Evaluation}\label{Subject-level Evaluation}
	 	To explore the effectiveness on the inter-trial variability, the introduced approach has been validated on 10 subjects from $S_1$ to $S_{10}$. For each evaluation, the total dataset was 53,760 samples. The results were given in \reftab{Subject-level-table}. 
	 	
	 	\begin{table}[h]
	 		\caption{Subject-level adaptation results for 10 individual subjects.}
	 		\label{Subject-level-table}
	 		\resizebox{\textwidth}{!}{
	 			\begin{tabular}{cccccccccccc}
	 				\hline
	 				\textbf{No. of Subjects} & \textbf{1} & \textbf{2} & \textbf{3} & \textbf{4} & \textbf{5} & \textbf{6} & \textbf{7} & \textbf{8} & \textbf{9} & \textbf{10} & \textbf{Average} \\ \hline
	 				Accuracy(\%)   & 97.49 & 87.69 & 96.73 & 97.01 & 92.89 & 90.70 & 88.08 & 96.84 & 96.32 & 98.08 & 94.18   \\
	 				F1 Score (\%)   & 97.50 & 87.55 & 96.78 & 97.05 & 92.86 & 90.62 & 87.97 & 96.89 & 96.39 & 98.09 & 94.17 \\ \hline 
	 			\end{tabular}
	 		}
	 	\end{table}
	 	
	 	Evaluated on the individual participant, the high of the accuracy was 98.08\% and the average was 94.18\% via the subject level analysis. Macro-averaged F1 score was also used to measure the performance of the presented model. The average of F1 score was 94.17\%, and the high was 98.09\%. The subject-level performance has indicated that the introduced approach was able to be applied to a single subject by precious prediction of MI tasks. It paved the road to build real-world EEG MI-based applications at the subject level, e.g., clinical translation of the technology. Meanwhile, it was suggested that the graph-based EEG analysis may become one of the research focus in the field of neuroscience.  
		
		\subsection{Group-wise Cross-validation}\label{Group-wise Cross-validation}
		\begin{figure}[h]
			\centering
			\begin{minipage}[t]{.48\linewidth}
				\includegraphics[width=2.5in]{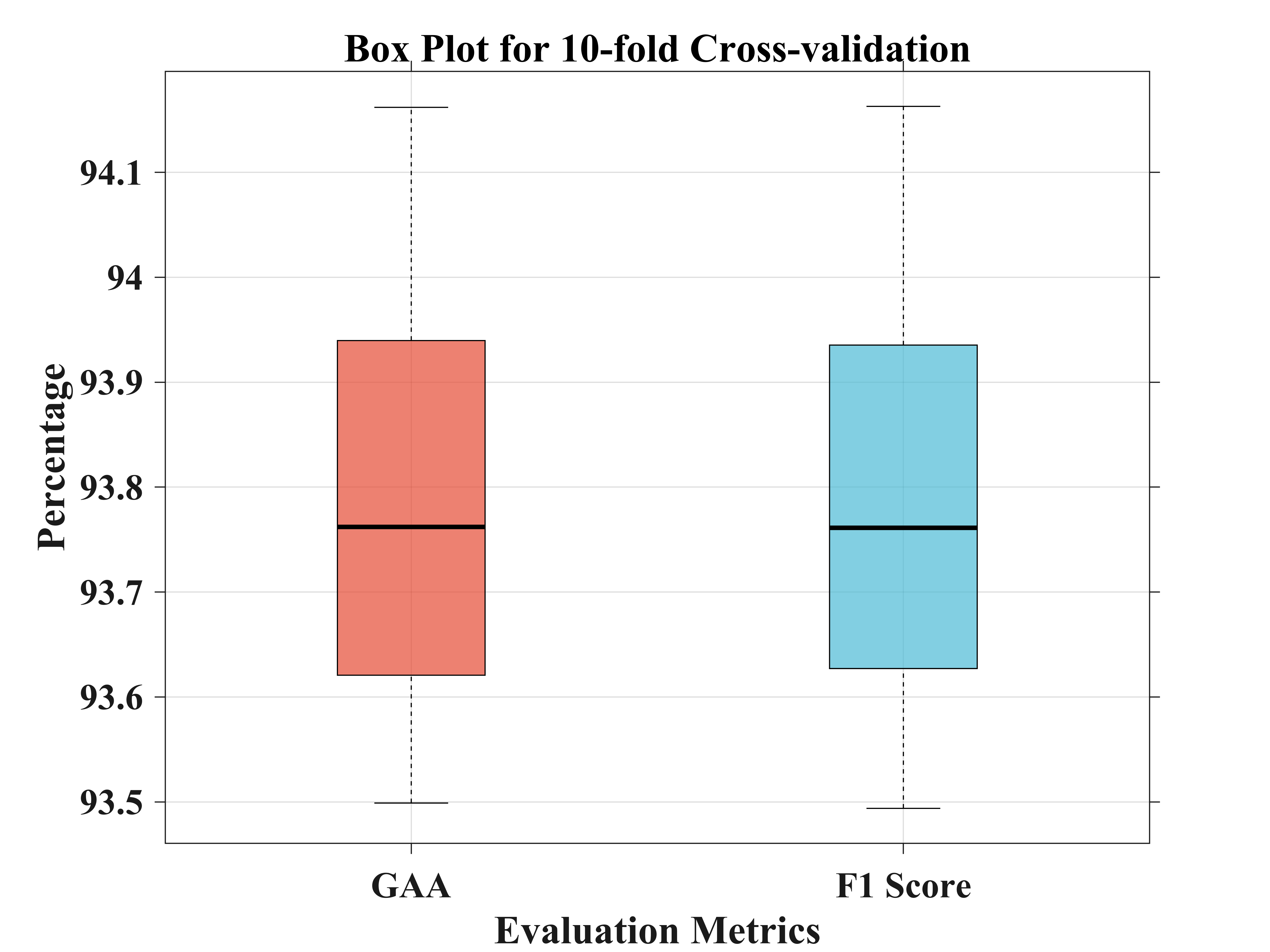}
				\subcaption{Box Plot for 10-fold Cross-validation}
				\label{Box Plot}
			\end{minipage}
			\begin{minipage}[t]{.48\linewidth}
				\includegraphics[width=2.17in]{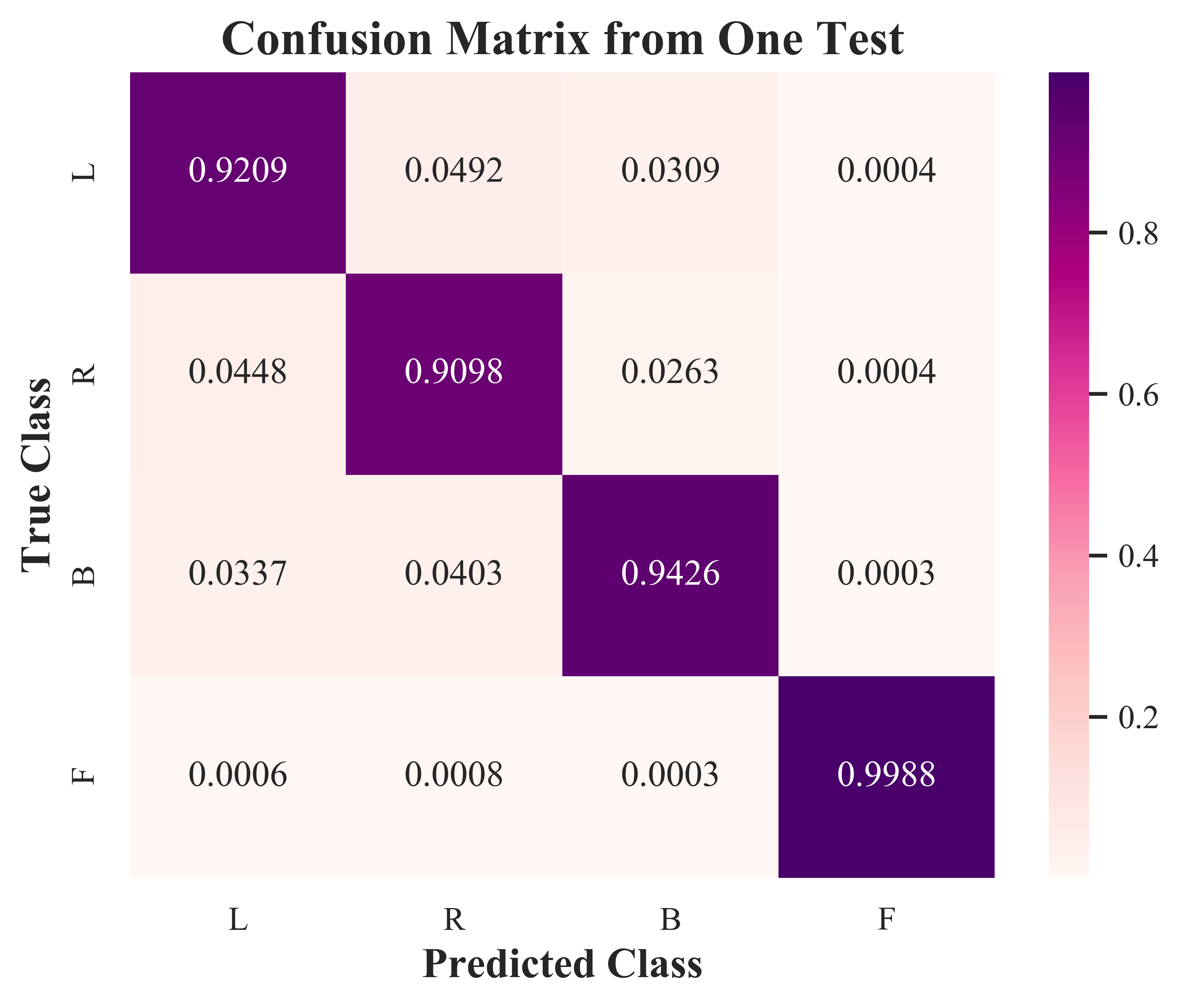}
				\subcaption{Confusion Matrix from One Individual Test}
				\label{Confusion Matrix}
			\end{minipage}
			\caption{Box plot for 10-fold cross-validation, and the confusion matrix from one individual test.}
			\label{Box Plot for 10-fold Cross-validation}
		\end{figure}
		
		Next, we 10-fold cross-validated the proposed model 2. The dataset we used was from 20 subjects, which was the same as that in \refsec{Model Comparison}. The box plot was demonstrated in \reffig{Box Plot for 10-fold Cross-validation}. The median accuracy was 93.76\%, and the high was 94.16\%. Besides, the average was 93.78\%, which was the state-of-the-art for group-wise EEG analysis among current studies. From the repetitive experiments, the stability and reliability of our presented model have been verified, in which the results can be stably reproduced. The reason under the outcome was that the proposed approach took consideration of the topological relationship of EEG electrodes, and made full advantage of network connectivity of brain dynamics. 
		
		Further, as shown in \reffig{Accuracy_Model_Comparison}, the Proposed Model 2 has achieved 94.28\% accuracy. It has been evaluated to obtain separate accuracy on each task. Displayed in \reffig{Confusion Matrix}, the separate accuracy on task L, R, B, F were 92.09\%, 90.98\%, 94.26\%, and 99.88\%. The data indicated that the introduced approach was superior for the prediction of imaging both feet. It was suggested that the method managed to tackle the concern of inter-subject variability due to its robustness and effectiveness. Plus, the group-specific model can be shared with a group of subjects with highly accurate prediction. Thus, it may pave the road towards building an universal EEG MI model for both neuroscience research and practical clinical translation. 
		
		\subsection{Comparison with State of the Art}\label{Comparison with State of the Art}
		The proposed approach has been carried out to compare the performance with current competitive models on the EEG Motor Movement/Imagery Dataset in \reftab{Performance Comparison with State-of-the-art}.
		
			\begin{table}[h]
			\centering
			\caption{Performance comparison with state-of-the-art}
			\label{Performance Comparison with State-of-the-art}
			\begin{tabular}{ccc}
				\hline
				\textbf{Approach} & \textbf{Max Accuracy} & \textbf{Num of Subjects} \\ \hline
				CNNs (2018)\cite{dose2018end} & 68.51\% & 1 \\ \hline
				RNNs (2018)\cite{ma2018improving} & 68.20\% & 12 \\ \hline
				Attention-based CNNs (2019)\cite{zhang2019graph} & 76.36\% & 1 \\ \hline
				\multirow{2}{*}{ESI+CNNs (2020)\cite{hou2020novel}} & 96.00\% & 1 \\ 
				& 92.50\% & 14 \\ \hline
				\multirow{2}{*}{\begin{tabular}[c]{@{}c@{}}\textbf{Attention-based}\\ \textbf{graph ResNet}\end{tabular}} & \textbf{98.08\%} & \textbf{1} \\ 
				& \textbf{94.28\%} & \textbf{20} \\ \hline
			\end{tabular}
		\end{table}
	
		Hou \emph{et al.} put forward a Convolutional Neural Network (CNN) based approach to classify four-class EEG MI, which was only been implemented to 14 subjects\cite{hou2020novel}. Our method accomplished better performance ($+$1.78\%) using the data from more subjects. Zhang \emph{et al.} introduced an Attention-based CNNs, and achieved an average of 76.36\% accuracy for subject-independent prediction\cite{zhang2019graph}. However, there was still space to improve the performance for practical implementation in the real world. Although the work from Dose \emph{et al.} work has not obtained convincing results, but the transfer learning they presented was worth to exploit in our future work for the clinical translation of EEG-based BCI technologies\cite{dose2018end}. According to \reftab{Performance Comparison with State-of-the-art}, the evidence suggested that our model achieved predominant and promising results from the perspective of both subject and group levels. This may be reasonable to suppose that the Attention-based graph ResNet was superior to decode graph-structured EEG signals by considering the topological relationship of EEG electrodes. 
	
	\section{Conclusion}\label{Conclusion and Future Work}
	In this paper, we proposed a novel structure of the GCN to predict four-class MI tasks from raw EEG data, which took consideration of the topological relationship of EEG electrodes. The presented Attention-based graph ResNet was superior to decode EEG signals that entailed a novel deep learning approach intending to address the universal EEG-based challenges in neuroscience, and pave the road towards clinical translation of the technology. Dominant results have shown that the approach managed to handle the inter-subject and inter-trial variations in raw EEG data. In the future, we will exploit the model by representing the topological structure of EEG electrodes from 2D to 3D, and implementing the Attention-based graph ResNet from the 3D perspective.
	
	\bibliographystyle{splncs04}
	\bibliography{bibliography}

\begin{thebibliography}{10}
\providecommand{\url}[1]{\texttt{#1}}
\providecommand{\urlprefix}{URL }
\providecommand{\doi}[1]{https://doi.org/#1}

\bibitem{biasiucci2018brain}
Biasiucci, A., Leeb, R., Iturrate, I., Perdikis, S., Al-Khodairy, A., Corbet,
  T., Schnider, A., Schmidlin, T., Zhang, H., Bassolino, M., et~al.:
  Brain-actuated functional electrical stimulation elicits lasting arm motor
  recovery after stroke. Nature communications  \textbf{9}(1),  1--13 (2018)

\bibitem{defferrard2016convolutional}
Defferrard, M., Bresson, X., Vandergheynst, P.: Convolutional neural networks
  on graphs with fast localized spectral filtering. In: Advances in neural
  information processing systems. pp. 3844--3852 (2016)

\bibitem{dhillon2007weighted}
Dhillon, I.S., Guan, Y., Kulis, B.: Weighted graph cuts without eigenvectors a
  multilevel approach. IEEE transactions on pattern analysis and machine
  intelligence  \textbf{29}(11),  1944--1957 (2007)

\bibitem{dose2018end}
Dose, H., M{\o}ller, J.S., Iversen, H.K., Puthusserypady, S.: An end-to-end
  deep learning approach to mi-eeg signal classification for bcis. Expert
  Systems with Applications  \textbf{114},  532--542 (2018)

\bibitem{goldberger2000physiobank}
Goldberger, A.L., Amaral, L.A., Glass, L., Hausdorff, J.M., Ivanov, P.C., Mark,
  R.G., et~al.: Physiobank, physiotoolkit, and physionet: components of a new
  research resource for complex physiologic signals. Circulation
  \textbf{101}(23),  e215--e220 (2000)

\bibitem{he2016deep}
He, K., Zhang, X., Ren, S., Sun, J.: Deep residual learning for image
  recognition. In: Proceedings of the IEEE conference on computer vision and
  pattern recognition. pp. 770--778 (2016)

\bibitem{hou2020novel}
Hou, Y., Zhou, L., Jia, S., Lun, X.: A novel approach of decoding eeg
  four-class motor imagery tasks via scout esi and cnn. Journal of neural
  engineering  \textbf{17}(1),  016048 (2020)

\bibitem{li2019deepgcns}
Li, G., Muller, M., Thabet, A., Ghanem, B.: Deepgcns: Can gcns go as deep as
  cnns? In: Proceedings of the IEEE International Conference on Computer
  Vision. pp. 9267--9276 (2019)

\bibitem{lv2014holistic}
Lv, J., Jiang, X., Li, X., Zhu, D., Zhang, S., Zhao, S., Chen, H., Zhang, T.,
  Hu, X., Han, J., et~al.: Holistic atlases of functional networks and
  interactions reveal reciprocal organizational architecture of cortical
  function. IEEE Transactions on Biomedical Engineering  \textbf{62}(4),
  1120--1131 (2014)

\bibitem{lv2017n}
Lv, J., Nguyen, V.T., van~der Meer, J., Breakspear, M., Guo, C.C.: N-way
  decomposition: Towards linking concurrent eeg and fmri analysis during
  natural stimulus. In: International Conference on Medical Image Computing and
  Computer-Assisted Intervention. pp. 382--389. Springer (2017)

\bibitem{ma2018improving}
Ma, X., Qiu, S., Du, C., Xing, J., He, H.: Improving eeg-based motor imagery
  classification via spatial and temporal recurrent neural networks. In: 2018
  40th Annual International Conference of the IEEE Engineering in Medicine and
  Biology Society (EMBC). pp. 1903--1906. IEEE (2018)

\bibitem{mahmood2019fully}
Mahmood, M., Mzurikwao, D., Kim, Y.S., Lee, Y., Mishra, S., Herbert, R.,
  Duarte, A., Ang, C.S., Yeo, W.H.: Fully portable and wireless universal
  brain--machine interfaces enabled by flexible scalp electronics and deep
  learning algorithm. Nature Machine Intelligence  \textbf{1}(9),  412--422
  (2019)

\bibitem{shanechi2019brain}
Shanechi, M.M.: Brain--machine interfaces from motor to mood. Nature
  neuroscience  \textbf{22}(10),  1554--1564 (2019)

\bibitem{shuman2013emerging}
Shuman, D.I., Narang, S.K., Frossard, P., Ortega, A., Vandergheynst, P.: The
  emerging field of signal processing on graphs: Extending high-dimensional
  data analysis to networks and other irregular domains. IEEE signal processing
  magazine  \textbf{30}(3),  83--98 (2013)

\bibitem{song2018eeg}
Song, T., Zheng, W., Song, P., Cui, Z.: Eeg emotion recognition using dynamical
  graph convolutional neural networks. IEEE Transactions on Affective Computing
   (2018)

\bibitem{vaswani2017attention}
Vaswani, A., Shazeer, N., Parmar, N., Uszkoreit, J., Jones, L., Gomez, A.N.,
  Kaiser, {\L}., Polosukhin, I.: Attention is all you need. In: Advances in
  neural information processing systems. pp. 5998--6008 (2017)

\bibitem{vidaurre2017brain}
Vidaurre, D., Smith, S.M., Woolrich, M.W.: Brain network dynamics are
  hierarchically organized in time. Proceedings of the National Academy of
  Sciences  \textbf{114}(48),  12827--12832 (2017)

\bibitem{wang2018eeg}
Wang, X.h., Zhang, T., Xu, X.m., Chen, L., Xing, X.f., Chen, C.P.: Eeg emotion
  recognition using dynamical graph convolutional neural networks and broad
  learning system. In: 2018 IEEE International Conference on Bioinformatics and
  Biomedicine (BIBM). pp. 1240--1244. IEEE (2018)

\bibitem{yang2016hierarchical}
Yang, Z., Yang, D., Dyer, C., He, X., Smola, A., Hovy, E.: Hierarchical
  attention networks for document classification. In: Proceedings of the 2016
  conference of the North American chapter of the association for computational
  linguistics: human language technologies. pp. 1480--1489 (2016)

\bibitem{zhang2019graph}
Zhang, D., Yao, L., Chen, K., Wang, S., Haghighi, P.D., Sullivan, C.: A
  graph-based hierarchical attention model for movement intention detection
  from eeg signals. IEEE Transactions on Neural Systems and Rehabilitation
  Engineering  \textbf{27}(11),  2247--2253 (2019)

\bibitem{zhang2019gcb}
Zhang, T., Wang, X., Xu, X., Chen, C.P.: Gcb-net: Graph convolutional broad
  network and its application in emotion recognition. IEEE Transactions on
  Affective Computing  (2019)

\end{thebibliography}

\end{document}